\documentclass[10pt, a4paper]{article}
\usepackage {fullpage}						
\usepackage {ucs}
\usepackage [utf8x]{inputenc}
\usepackage {graphicx}
\usepackage [bookmarks=false, pdfstartview={FitH}, colorlinks=true, linkcolor=blue, citecolor=blue, urlcolor=blue]{hyperref} 
\usepackage {amsfonts,amsmath,amsthm,amssymb}
\usepackage {latexsym}
\usepackage {graphics, graphicx}
\usepackage {fancyhdr}
\usepackage[table]{xcolor}
\usepackage {epstopdf} 						
\usepackage {multicol}						
\usepackage {multirow}  					
\usepackage {soul} 							
\usepackage {authblk} 						
\usepackage {datetime} 						
\usepackage {subfigure}						
\usepackage {parskip}						
\usepackage {setspace}						
\usepackage[font=small,textfont=it,width=0.97\textwidth]{caption}		
\usepackage{mathtools}						
\usepackage{txfonts}						

\newcommand {\pd} {\partial}
\newcommand {\R} {{\rm I\!R}}

\renewcommand {\P} {{\rm I\!P}}
\newcommand {\umu} {{\muup}}

\renewcommand{\(}{\left(}
\renewcommand{\)}{\right)}
\renewcommand {\a} {{\alpha}}
\newcommand {\nn}{\nonumber}
\renewcommand{\vec}{\mathbf}

    \newtheorem{lemma}{Lemma}[section]
\theoremstyle{definition}   
    \newtheorem{definition}{Definition}[section]
\theoremstyle{remark}       
    \newtheorem{remark}{Remark}[section]

\def\subtext#1{{\mbox{\scriptsize #1}}}

\newcommand {\startenv} {~~\\ \begin{tabular}{||l}\parbox[t]{0.95\linewidth}}
\newcommand {\stopenv} 	{\end{tabular} \\~~}

\title{
	Numerical treatment of the Filament Based Lamellipodium Model (FBLM)
	}
	
\author{
	A. Manhart \thanks{Faculty of Mathematics, University of Vienna, Austria, ~\hfill {\tt angelika.manhart@univie.ac.at}}~{},
	D. Oelz \thanks{Courant Institute of Mathematical Sciences, New York University, USA , ~\hfill {\tt dietmar@cims.nyu.edu}}~{},
	C. Schmeiser \thanks{Faculty of Mathematics, University of Vienna, Austria, ~\hfill {\tt christian.schmeiser@univie.ac.at}}~{},
	N. Sfakianakis \thanks{Johannes-Gutenberg University of Mainz, Germany, ~\hfill {\tt sfakiana@uni-mainz.de}}~{},
	}
\date{}

\begin{document}
\maketitle

\begin{abstract}
	We describe in this work the numerical treatment of the Filament Based Lamellipodium Model (FBLM). The model itself is a two-phase two-dimensional continuum model, describing the dynamics of two interacting families of locally parallel F-actin filaments. It includes, among others, the bending stiffness of the filaments, adhesion to the substrate, and the cross-links connecting the two families. The numerical method proposed is a Finite Element Method (FEM) developed specifically for the needs of these problem. It is comprised of composite Lagrange-Hermite two dimensional elements defined over two dimensional space. We present some elements of the FEM and emphasise in the numerical treatment of the more complex terms. We also present novel numerical simulations and compare to in-vitro experiments of moving cells.
\end{abstract}



\section{Introduction}
The lamellipodium is a flat cell protrusion mechanism functioning as a motility organelle in protrusive cell migration \cite{Vinzenz_Small_Winkler2012}. It is a very dynamic structure mainly consisting of a network of branched actin filaments. 
These are semi-elastic rods that represent the polymer form of the protein actin. They are continuously remodeled by polymerization and depolymerization and therefore undergo treadmilling \cite{Blanchoin_Plastino2014}. Actin associated cross-linker proteins and myosin motor proteins integrate them into the lamellipodial meshwork which plays a key role in cell shape stabilization and in cell migration.
Different modes of cell migration result from the interplay of protrusive forces due to polymerization, actomyosin dependent contractile forces and regulation of cell adhesion \cite{LaemmermannSixt2009}.

The first modeling attempts have resolved the interplay of protrusion at the front and retraction at the rear in a one-dimensional spatial setting \cite{alt_dembo,gracheva_othmer}. Two-dimensional continuum models were developed in order to include the lateral flow of F-actin along the leading edge of the cell into the quantitative picture. Those models can explain characteristic shapes of amoeboid cell migration \cite{mogilner_rubinstein, Rubinstein_Mogilner_viscoelastic} on two-dimensional surfaces as well as the transition to mesenchymal migration \cite{Sakamoto2014}.

One of the still unresolved scientific questions concerns the interplay between macroscopic observables of cell migration and the microstructure of the lamellipodium meshwork. Specialised models had been developed separately from the continuum models to track microscopic information on filament directions and branching structure \cite{maly_borisy, schaus_taylor_borisy, lacayo_e_a}. However, solving fluid-type models that describe the whole cytoplasm while retaining some information on the microstructure of the meshwork has turned out to be challenging. One approach is to formulate hybrid models \cite{MetzkeMofrad_2009}, another one to directly formulate models on the computational, discrete level \cite{maree_ea}. 
Recently the approach to directly formulate a computational model has been even extended into the three-dimensional setting making use of a finite element discretization \cite{Mousavi2015}.

In an attempt to create a simulation framework that addresses the interplay of macroscopic features of cell migration and the meshwork structure the Filament Based Lamellipodium Model (FBLM) has been developed. It is a two-dimensional, two-phase, anisotropic continuum model for the dynamics of the lamellipodium network which retains key directional information on the filamenteous substructure of this meshwork \cite{Oelz2008}.

The model has been derived from a microscopic description based on the dynamics and interaction of individual filaments \cite{Oelz2010a}, and it has by recent extensions \cite{MSOS} reached a certain state of maturity. Since the model can be written in the form of a generalized gradient flow, numerical methods based on optimization techniques have been developed \cite{OelzSch-JMB,Oelz2008}. 
Numerical efficiency had been a shortcoming of this approach. This has led to the development of a Finite Element numerical method which is presented in this article alongside simulations of a series of migration assays. 


\section{Mathematical modeling}
\begin{figure}[t]
\begin{center}
	\includegraphics[width=0.8\linewidth]{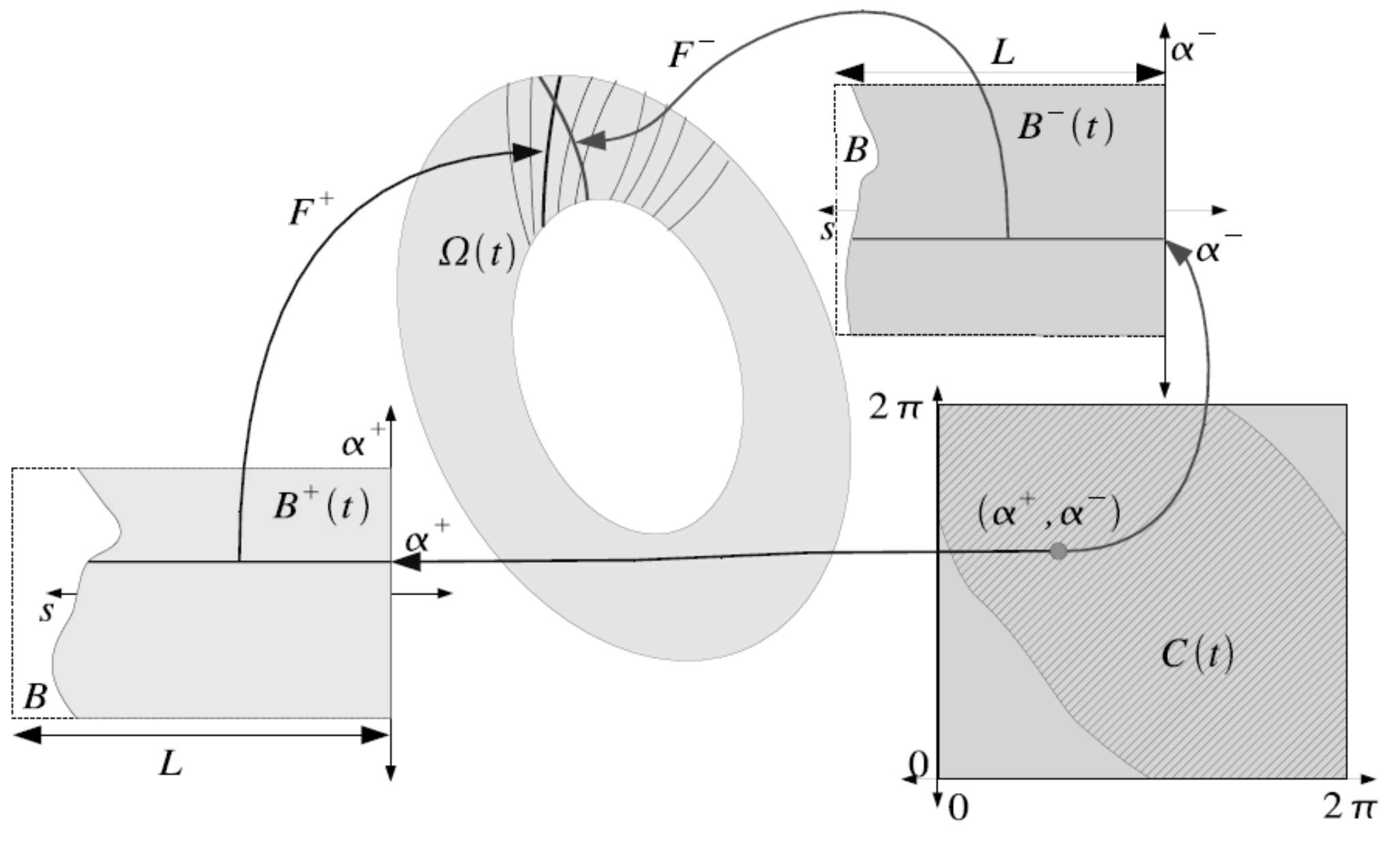}
\end{center}
\caption{Graphical representation of (\ref{eq:strong}); showing here the lamellipodium $\Omega(t)$ ``produced'' by the mappings $\vec{F}^\pm$ and the crossing--filament domain $\mathcal C$.}\label{fig:domains}
\end{figure}

In this section the FBLM will be sketched. For more detail see \cite{MSOS}.
The main unknowns of the model are the positions of the actin filaments in two locally
parallel families (denoted by the superscripts +  and -). Each of these families covers
a topological ring with all individual filaments connecting the inner boundary with the
outer boundary. The outer boundaries are the physical leading edge and therefore
identical, whereas the inner boundaries of the two families are artificial and may be
different. Filaments are labelled by $\alpha\in [0,2\pi)$, where the interval
represents a one-dimensional torus. The maximal arclength of the filaments 
in an infinitesimal element $d\alpha$ 
of the $\pm$-family at time $t$ is denoted by $L^\pm(\alpha,t)$, and an arclength 
parametrization of the filaments is denoted by 
$\left\{\vec{F}^\pm(\alpha,s,t): -L^\pm(\alpha,t)\le s \le 0\right\}\subset {\R}^2$, where the
leading edge corresponds to $s=0$, i.e.
\begin{equation}\label{tether}
  \left\{\vec{F}^+(\alpha,0,t): 0\le \alpha < 2\pi\right\} = \left\{\vec{F}^-(\alpha,0,t): 0\le \alpha < 2\pi\right\} 
  \qquad \forall\, t \;,
\end{equation}
which together with 
\begin{equation}\label{inext}
  \left|\partial_s \vec{F}^\pm(\alpha,s,t)\right| = 1 \qquad\forall\, (\alpha,s,t) \;,
\end{equation}
constitutes constraints for the unknowns $\vec{F}^\pm$. The second constraint is connected 
to an \emph{inextensibility} assumption on the filaments, which implies that $s$ can
also be interpreted as a monomer counter along filaments.

We expect that different filaments
of the same family do not intersect each other, and each plus-filament crosses each
minus-filament at most once. The first condition is guaranteed by 
det$(\partial_\alpha \vec{F}^\pm, \partial_s \vec{F}^\pm) >0$, where the sign indicates that
the labelling with increasing $\alpha$ is in the clockwise direction. The second 
condition uniquely defines $s^\pm = s^\pm(\alpha^+,\alpha^-,t)$ such that
$\vec{F}^+(\alpha^+,s^+,t) = \vec{F}^-(\alpha^-,s^-,t)$, for all
$(\alpha^+,\alpha^-) \in {\mathcal C}(t)$, the set of all pairs of crossing filaments.
As a consequence, there are coordinate transformations 
$\psi^\pm:\, (\alpha^\mp,s^\mp) \mapsto (\alpha^\pm,s^\pm)$ such that 
\[
  \vec{F}^\mp = \vec{F}^\pm \circ \psi^\pm \;.
\]
In the following, we shall concentrate on one of the two families and skip the 
superscripts except that the other family is indicated by the superscript $*$.
The heart of the FBLM is the force balance
\begin{align}
0=\underbrace{\mu^B \partial_s^2\left(\eta \partial_s^2 \vec{F}\right)}_\subtext{bending}
		&-\underbrace{\partial_s\left(\eta \lambda_{\rm inext} \partial_s \vec{F}\right)}_\subtext{in-extensibility}
		+\underbrace{\mu^A \eta D_t \vec{F}}_\subtext{adhesion} + \underbrace{\partial_s\left( p(\rho) \partial_\alpha \vec{F}^{\perp}\right)
 -\partial_\alpha \left( p(\rho) \partial_s \vec{F}^{\perp}\right)}_\subtext{pressure} \nn \\
&\underbrace{\pm\partial_s\left(\eta\eta^* \widehat{\mu^T} (\phi-\phi_0)\partial_s \vec{F}^{\perp}\right)}_\subtext{twisting}  
 	+ \underbrace{\eta\eta^* \widehat{\mu^S}\left(D_t \vec{F} - D_t^* \vec{F}^*\right)}_\subtext{stretching} \;,\label{eq:strong}
\end{align}
where the notation $\vec{F}^\bot = (F_1,F_2)^\bot = (-F_2,F_1)$ has been. For fixed $s$ and $t$, the function 
$\eta(\alpha,s,t)$, is the number density of
filaments of length at least $-s$ at time $t$ with respect to $\alpha$. Its dynamics
and that of the maximal length $L(\alpha,t)$ will not be discussed here. It can be 
modeled by incorporating the effects of polymerization, depolymerization, branching,
and capping (see \cite{MSOS}). We only note that, faster polymerization (even locally) leads to wider lamellipodia.

The first term on the right hand side of (\ref{eq:strong}) describes the
filaments$^\prime$ resistance against bending with the stiffness parameter 
$\mu^B>0$. The second term is a tangential tension force, which arises
from incorporating the inextensibility constraint (\ref{inext}) with the Lagrange
multiplier $\lambda_{\rm inext}(\alpha,s,t)$. The third term describes friction of the
filament network with the nonmoving substrate (see \cite{Oelz2010a} for its derivation 
as a macroscopic limit of the dynamics of transient elastic adhesion linkages). 
Since filaments
polymerize at the leading edge with the polymerization speed $v(\alpha,t)\ge 0$,
they are continuously pushed into the cell with that speed, and the material derivative
$$
  D_t \vec{F} := \partial_t \vec{F} - v\partial_s \vec{F}
$$
is the velocity of the actin material relative to the substrate. For the modeling of $v$
see \cite{MSOS}.

The second line of (\ref{eq:strong}) models a pressure effect caused by Coulomb
repulsion between neighboring filaments of the same family with pressure $p(\rho)$,
where the actin density in physical space is given by 
\begin{equation} \label{eq:rho}
    \rho = \frac{\eta}{\det(\partial_\alpha \vec{F}, \partial_s \vec{F})}  \;.
\end{equation}

Finally, the third line of (\ref{eq:strong}) models the interaction between the two
families caused by transient elastic cross-links and/or branch junctions. The first 
term describes elastic resistance against changing the angle 
$\phi=\arccos (\partial_s \vec{F}\cdot\partial_s \vec{F}^*)$ between filaments away from the 
equilibrium angle $\phi_0$ of the equilbrium conformation of the cross-linking 
molecule. The last term describes friction between the two families analogously to
 friction with the substrate. The friction coefficients have the form
 $$
    \widehat{\mu^{T,S}} = \mu^{T,S} \left| \frac{\partial\alpha^*}{\partial s} \right| \;,
 $$
 with $\mu^{T,S}>0$ and the partial derivative refers to the coordinate transformation
 $\psi^*$, which is also used when evaluating partial derivatives of $\vec{F}^*$.
 
The system (\ref{eq:strong}) is considered subject to the boundary conditions 
\begin{align} \label{eq:newBC}
	- \mu^B\partial_s\left(\eta\partial_s^2 \vec{F}\right) - p(\rho)\partial_\alpha \vec{F}^\perp 
		&+ \eta \lambda_{\rm inext} \partial_s \vec{F}  
	   \mp\eta\eta^* \widehat{\mu^T}(\phi-\phi_0)\partial_s \vec{F}^\perp \\
	&=\left\{
		\begin{array}{l l}
			\eta \left(f_{\rm tan}(\alpha)\partial_s \vec{F} + f_{\rm inn}(\alpha) \vec{V}(\alpha)\right), & \quad \mbox{for 	} s=-L \;,\\
			 \pm\lambda_{\rm tether} \nu, & \quad  \mbox{for }  s=0 \;,
		\end{array}\right.\nonumber\\
	\eta \partial_s^2 \vec{F} &= 0, \qquad \mbox{for } s=-L,0 \;.\nonumber
\end{align}

The terms in the second line are forces applied to the filament ends. The force in the 
direction $\nu$ orthogonal to the leading edge at $s=0$ arises from the constraint
(\ref{tether}) with the Lagrange parameter $\lambda_{\rm tether}$. Its biological
interpretation is due to tethering of the filament ends to the leading edge. The forces
at the inner boundary $s=-L$ are models of the contraction effect of actin-myosin 
interaction in the interior region (see \cite{MSOS} for details).

\section{Numerical method}
We present in this section some elements of the Finite Element (FE) method that we have developed for (\ref{eq:strong}). We prefer in this work a more computational approach to highlight the implementation difficulties encountered, and to emphasize on the ``special'' treatment that some of the terms necessitate \cite{Braess.2001,FEM.MATLAB.2000}. We refer to \cite{MOSS} for further details and a thorough theoretical presentation of the FEM.

\subsection{Discrete formulation}
As previously, we skip the superscripts $(\pm)$ except for those of the other family that we indicate by $*$.

The domain $B$ of $\vec{F}$ (see also Fig. \ref{fig:domains}) is discretized into isodynamous computational cells of the form
\begin{equation}\label{eq:cell}
	C_{i,j}=[\a_i,\a_{i+1})\times[s_j,s_{j+1})\;,
\end{equation}
where $\a_{i+1}-\a_i=\Delta \a$ and $s_{i+1}-s_i=\Delta s$ for $i=1\dots N_\a-1$, and $j=1\dots N_s-1$. We denote by $\mathcal T_{\Delta \a,\Delta s}$ the discretization of $B$. Let now, $C$ represent the generic cell of $\mathcal T_{\Delta \a,\Delta s}$, i.e.
\begin{equation}\label{eq:C}
	C= [\a_1,\a_2)\times[s_1,s_2),\ \mbox{with }  \a_2-\a_1=\Delta \a >0, \  s_2-s_1=\Delta s>0\;,
\end{equation}
with its vertices in lexicographic order: 
\begin{equation}\label{eq:verC}
	V_1=(\a_1,s_1),\quad V_2=(\a_1,s_2),\quad  V_3=(\a_2,s_1),\quad V_4=(\a_2,s_2)\;.
\end{equation}

\begin{figure}[t]
	\centering 	
	\includegraphics[width=0.35\linewidth]{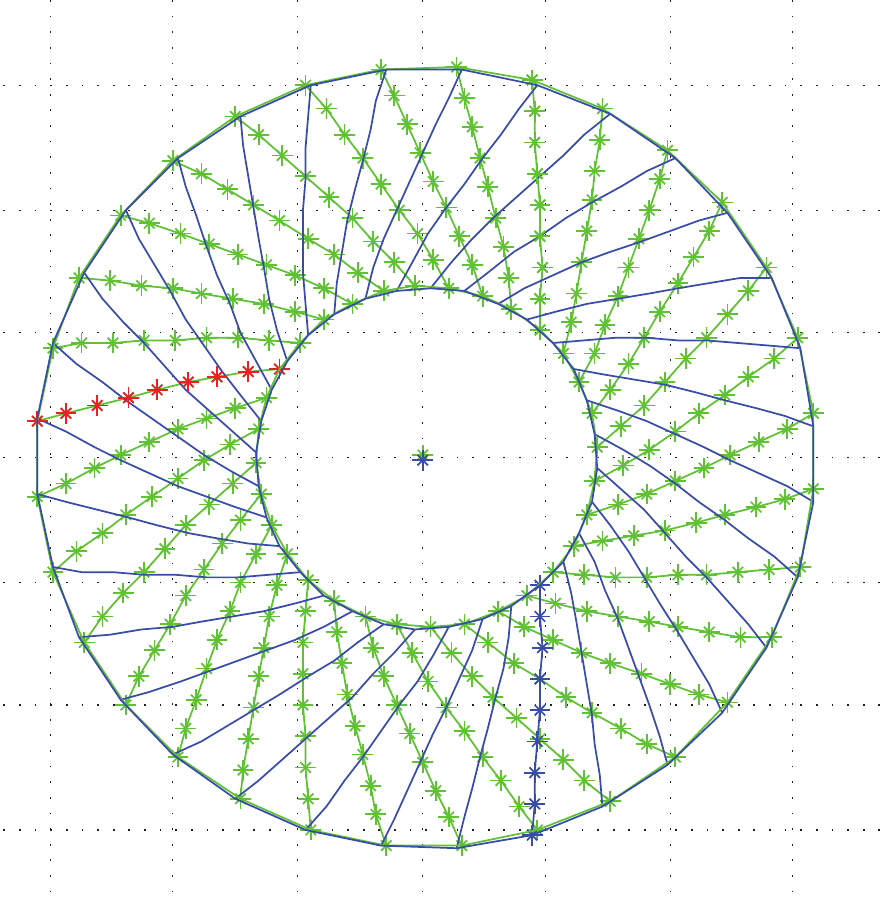}~~~
	\includegraphics[width=0.35\linewidth]{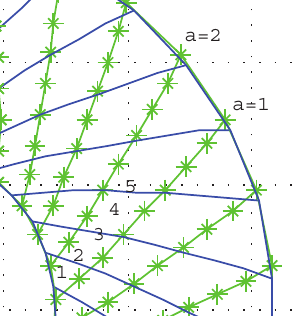} \\
	\caption{Discretized lamellipodium (left) and lamellipodium fragment (right).}\label{fig:discr}
\end{figure}

The composite \emph{Lagrange-Hermite scalar polynomial space} over the cell $C$ is defined as:
\begin{equation}\label{eq:Vc}
	\mathcal V_C=\Big\{v\in \P[\a,s], \ (\a,s)\in C \ \Big| \ v(\a,\cdot)\in\P_3^C[s]\mbox{ and } v(\cdot,s)\in\P_1^C[\a]  \Big\}\;,
\end{equation}
where $\P_k^C[\cdot]$ denotes the vector space of scalar real polynomials, with real coefficients, of a single variable in the corresponding component of $C$, and degree at most $k$.
\begin{remark}
	The higher smoothness of the elements of $\mathcal V_C$ along the $s$-direction is necessitated by the higher order $s$-derivatives of $\vec{F}$ in (\ref{eq:strong}).
\end{remark}

\begin{figure}[t]
	\centering
%

	\subfigure[$H_1^C$] {\includegraphics[width=0.22\textwidth]{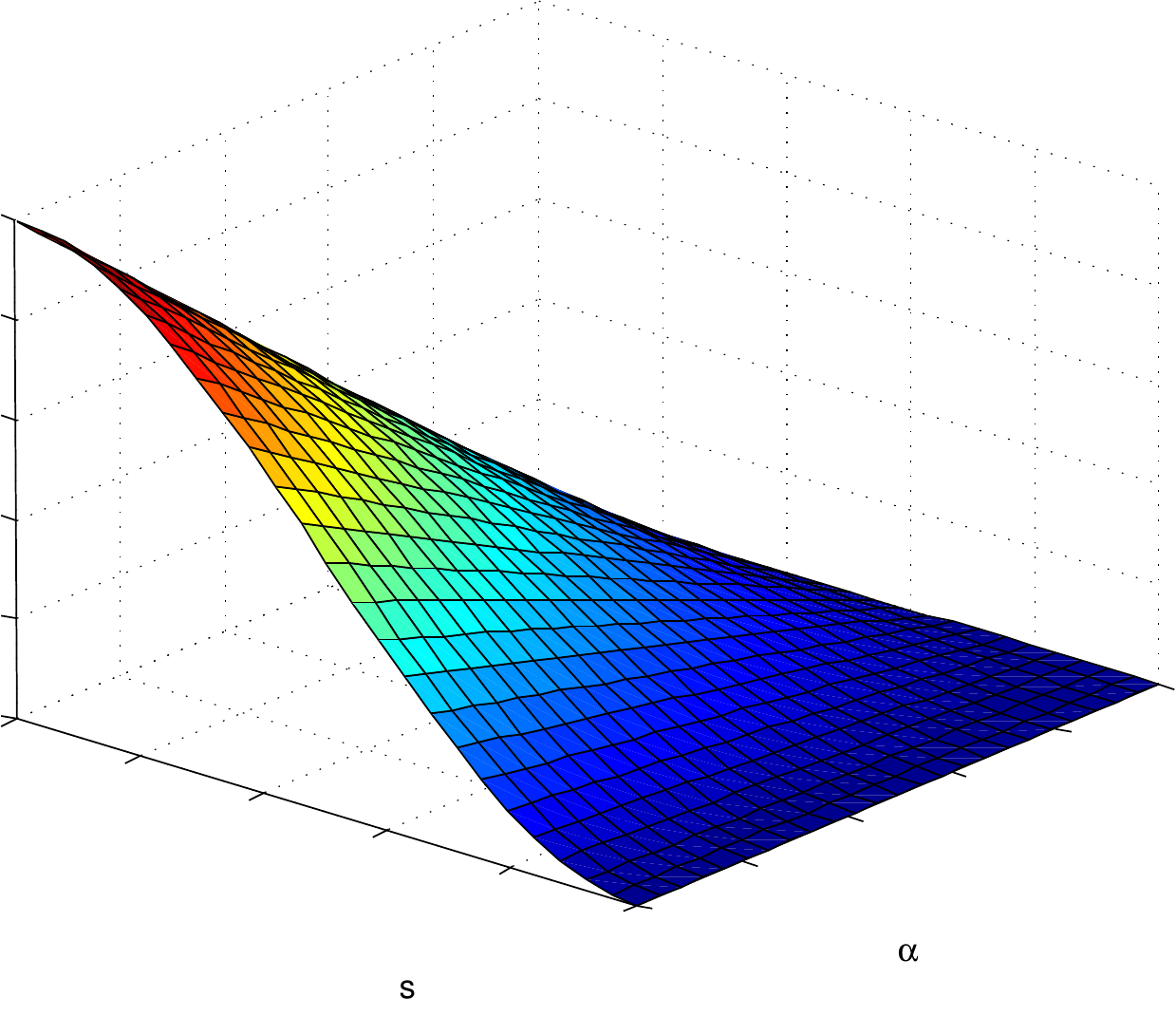}}~~
	\subfigure[$H_2^C$] {\includegraphics[width=0.22\textwidth]{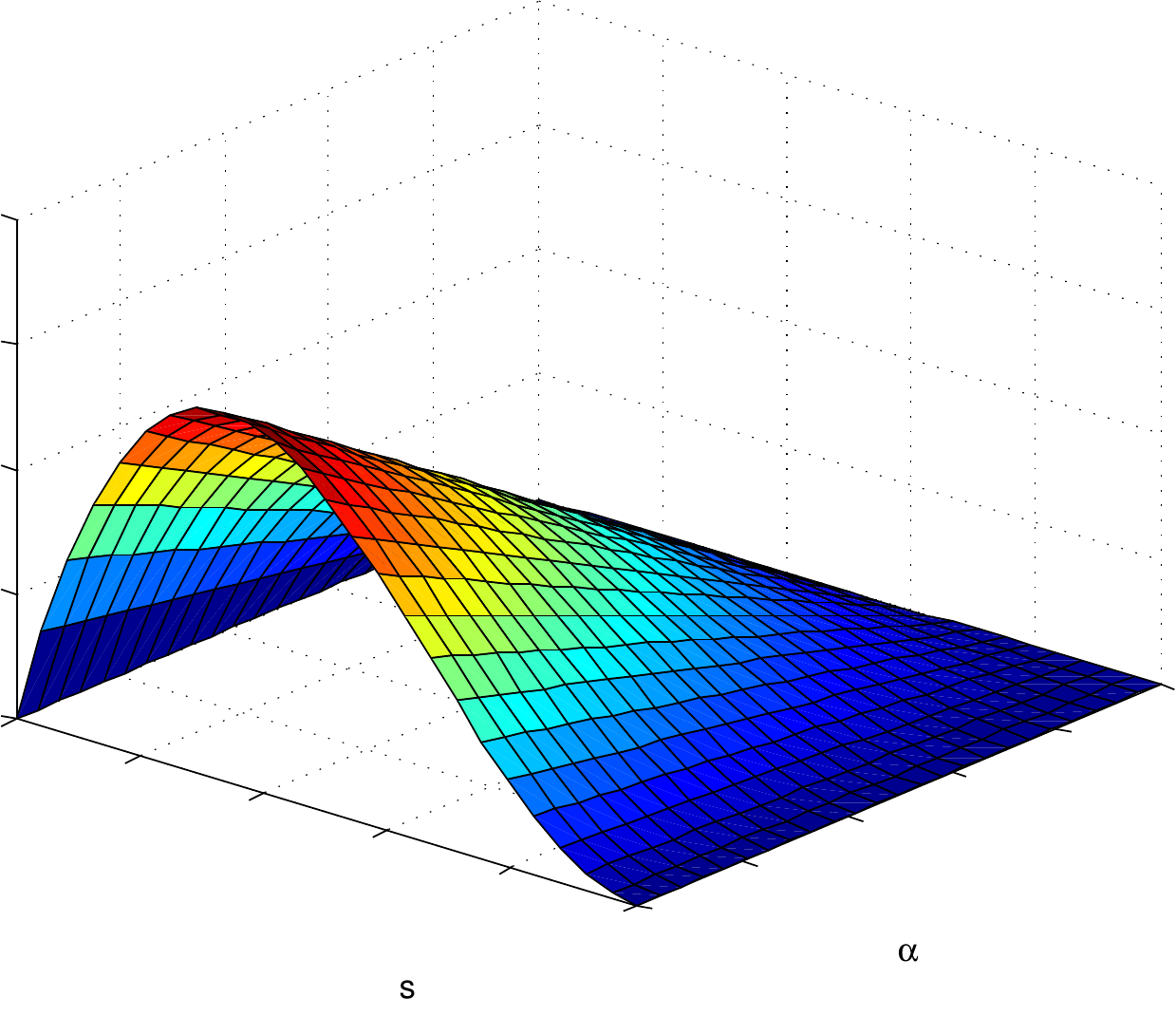}}~~
	\subfigure[$H_3^C$]	{\includegraphics[width=0.22\textwidth]{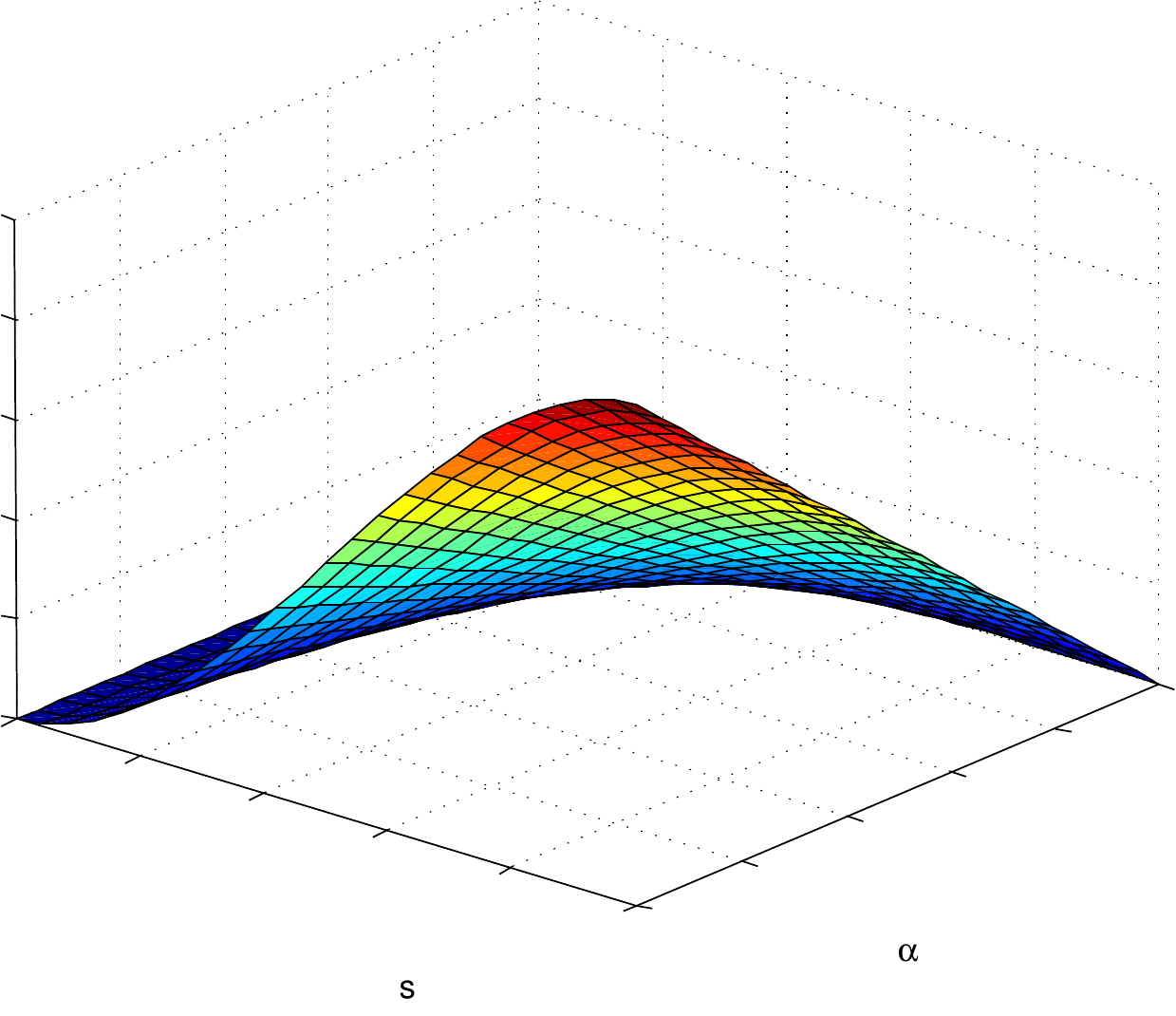}}~~
	\subfigure[$H_4^C$] {\includegraphics[width=0.22\textwidth]{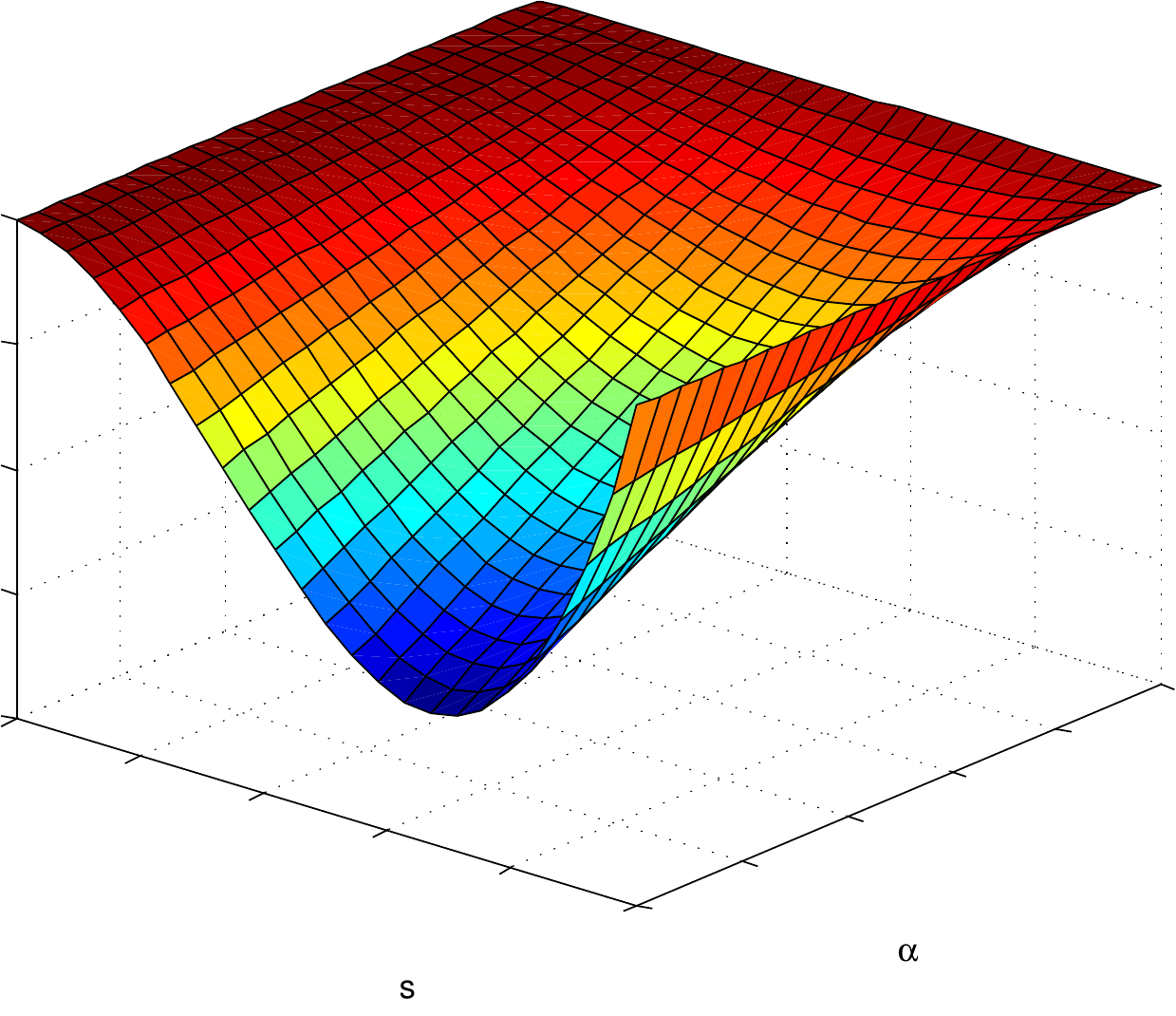}} 
	\caption{Graphical representation of the Lagrange-Hermite shape functions (\ref{eq:H}). Each one of the shape 
	functions attains the value 1 in one degree of freedom, and 0 on all the rest.}\label{fig:shape}
\end{figure}

We also set, for $(a,s)\in C$, the \emph{shape functions}: 
\begin{equation}\label{eq:LG}
\left.\begin{array}{lcl}
		L_1^C(\a) =\frac{\a_2-\a}{\Delta \a}, 									
				&~&\quad G_1^C(s)=1-\frac{3(s-s_1)^2}{\Delta s^2}+\frac{2(s-s_1)^3}{\Delta s^3}, \\
		L_2^C(\a)=1-L_1^C(\a),
				&&\quad G_2^C(s)=s-s_1-\frac{2(s-s_1)^2}{\Delta s}+\frac{(s-s_1)^3}{\Delta s^2},\\
				&&\quad G_3^C(s)=1-G_1^C(s),\\
				&&\quad G_4^C(s)=-G_2^C(s_1+s_2-s),
\end{array}\right\}
\end{equation}
which satisfy:
\begin{equation}
	\left.\begin{array}{lclclcl}
	L_1^C(\a_1)= 1,&&\quad L_1^C(\a_2)=0, \\
	L_2^C(\a_1)= 0,&&\quad	 L_2^C(\a_2)=1, \\
	G_1^C(s_1)= 1,&~&\quad	 G_1^C(s_2)=0,&~&\quad \frac{\pd}{\pd s}G_1(s_1)=0,&~&\quad \frac{\pd}{\pd s}G_1(s_2)=0,\\
	G_2^C(s_1)= 0,&&\quad	 G_2^C(s_2)=0,&&\quad \frac{\pd}{\pd s}G_2(s_1)=1,&&\quad \frac{\pd}{\pd s}G_2(s_2)=0,\\
	G_3^C(s_1)= 0,&&\quad	 G_3^C(s_2)=1,&&\quad \frac{\pd}{\pd s}G_3(s_1)=0,&&\quad \frac{\pd}{\pd s}G_3(s_2)=0,\\
	G_4^C(s_1)= 0,&&\quad	 G_4^C(s_2)=0,&&\quad \frac{\pd}{\pd s}G_4(s_1)=0,&&\quad \frac{\pd}{\pd s}G_4(s_2)=1.
	\end{array}\right\}
\end{equation}
and that they span $\P_1^C[\a]$ and $\P_3^C[s]$ respectively. Accordingly we set the composite \emph{Lagrange-Hermite shape functions}, for  $(\a,s)\in C$, as (see also \cite{Braess.2001}):
\begin{equation}\label{eq:H}
	\left.\begin{array}{lcl}
		H_1^C(\a,s)=L_1^C(\a) \, G_1^C(s),&~& H_5^C(\a,s)=L_2^C(\a) \, G_1^C(s),\\
		H_2^C(\a,s)=L_1^C(\a) \, G_2^C(s),&& H_6^C(\a,s)=L_2^C(\a) \, G_2^C(s),\\
		H_3^C(\a,s)=L_1^C(\a) \, G_3^C(s),&& H_7^C(\a,s)=L_2^C(\a) \, G_3^C(s),\\
		H_4^C(\a,s)=L_1^C(\a) \, G_4^C(s),&& H_8^C(\a,s)=L_2^C(\a) \, G_4^C(s), 
	\end{array}\right\}
\end{equation}
and for $(\a,s)\not\in C$ as $H_i^C(\a,s)=0$, $i=1\dots8$. Refer to Fig. \ref{fig:shape} for a graphical representation presentation of (\ref{eq:H}).

It is easy to verify the following Lemma:
\begin{lemma}\label{th:basis_2}
	Let $C$ given by (\ref{eq:C}) and $\mathcal V_C$ by (\ref{eq:Vc}). The following statements hold:
	\begin{description}
	\item[\textbullet] Every element $v\in \mathcal V_C$ is uniquely determined by the eight Degrees Of Freedom (DOFs): point values $v(V_i)$ and $s$-derivatives $\frac{\pd}{\pd s}v(V_i)$ at the four vertices $V_i, \ i=1\dots 4$ of $C$.
	\item[\textbullet] The Lagrange-Hermite shape functions (\ref{eq:H}) constitute a basis for $\mathcal V_C$.	
	\end{description}
\end{lemma}

We are able now to define the \emph{interpolation} over the cell $C$. In particular, let $v\in\mathcal  V_C$ and $c_i,d_i$ the DOFs corresponding to the point $v(V_i)$ and $s$-derivative values $\frac{\pd}{\pd s}v(V_i)$ of $v$ over the vertices of $C$. The \emph{scalar local interpolation function} $I^C(\cdot,\cdot)\in \mathcal V_C$ reads as:
	\begin{equation}\label{eq:IC}
		I^C(\a,s)=\sum_{i=1}^4 c_i\,H_{2i-1}^C(\a,s) + d_i\,H_{2i}^C(\a,s), \quad (\a,s)\in C\;.
	\end{equation}

Similarly, we define the \emph{scalar global interpolation function}, as a concatenation of the local interpolation functions $I^C(\cdot,\cdot)$,  for all $C\in\mathcal T_{\Delta \a,\Delta s}$. To that end, we first note that every internal node $(\a_i,s_j)$ of the discretization is a vertex to four computational cells, namely:
\begin{equation}\label{eq:neigh}
\left. \begin{array}{lcl}
	C_{i-1,j-1}	=[\a_{i-1},\a_i)\times[s_{j-1},s_j),&\qquad& C_{i-1,j}	=[\a_{i-1},\a_i)\times[s_j,s_{j+1}),\quad \\ 
	C_{i,j-1}	=[\a_i,\a_{i+1})\times[s_{j-1},s_j),&& C_{i,j}	=[\a_i,\a_{i+1})\times[s_j,s_{j+1}). 
\end{array}\right\}
\end{equation}
It is moreover equipped with two degrees of freedom, $v_{i,j}$ and $d_{i,j}$ that correspond to the point and the $s$-derivative values of the interpolated function respectively. In view of Lemma \ref{th:basis_2} the interpolation function can be uniquely (re-)constructed by the ``full'' set of DOFs.
\begin{definition}[Scalar global interpolation function]\label{def:I}
	Let
	\[ \Big\{v_{i,j},\; d_{i,j},\ i=1\dots N_\a,\; j=1\dots N_s\Big\} \]
	be a set of DOFs corresponding to point values and $s$-derivatives over all discretization nodes. The interpolation function to these values reads as:
	\begin{align}
		I(\a,s)	=&\sum_{i=1}^{N_\a} \sum_{j=1}^{N_s} 
					v_{i,j} \(H_7^{C_{i-1,j-1}} + H_5^{C_{i-1,j}} + H_3^{C_{i,j-1}} + H_1^{C_{i,j}}\)(\a,s)\nn\\
				+&\sum_{i=1}^{N_\a} \sum_{j=1}^{N_s}
					d_{i,j}\(H_8^{C_{i-1,j-1}} + H_6^{C_{i-1,j}} + H_4^{C_{i,j-1}} + H_2^{C_{i,j}}\)(\a,s)\;,\label{eq:I}
	\end{align}
	for $\ (\a,s)\in B$.
\end{definition}

Similarly, the functions that belong piecewise in $\mathcal V_C$ constitute the components of the \emph{scalar global approximation space}:
\begin{equation}\label{eq:V}
	\mathcal V=\left\{ v:B\longrightarrow \R\ \Big| \ v \big|_C \in \mathcal V_C, \ \forall C \in \mathcal T_{\Delta \a, \Delta s}  \right\}\;.
\end{equation}

We proceed to the 2 dimensional case and first introduce the \emph{2D Lagrange-Hermite polynomial space} as:
\begin{equation}
	\mathcal V_C^{2d}=\left\{ \(v_x\; ,\; v_y\)^\top \mbox{ with } v_x, v_y \in \mathcal V_C  \right\}\;,
\end{equation}
for $C$ given by (\ref{eq:C}) and $\mathcal V_C$ by (\ref{eq:Vc}). We note that $\mathcal V_C^{2d}$ is a linear space of dimension 16.

Given now the sixteen DOFs $c_i^{x,y}$, $d_i^{x,y}$, $i=1\dots 4$ that correspond to the point and $s$-derivative values of a function in $\mathcal V_C^{2d}$ the \emph{2D local interpolation function} reads as follows:
\begin{align}
	\textbf I^C(\a,s)	&=\left( \sum_{i=1}^4 \( c_i^x\,H_{2i-1}^C + d_i^x\,H_{2i}^C\) (\a,s)\; ,\;  
				\sum_{i=1}^4 \(c_i^y\,H_{2i-1}^C+ d_i^y\,H_{2i}^C\) (\a,s)\right)^\top \nn \\
						&=\sum_{i=1}^4 (c_i^x\; ,\; c_i^y)^\top H_{2i-1}^C(\a,s)+\sum_{i=1}^4 (d_i^x\; ,\; d_i^y)^\top H_{2i}^C(\a,s)\;. \label{eq:IC2D}
\end{align}	

Accordingly, follows the \emph{2D global interpolation function}, 
\begin{definition}[2D global interpolation function]\label{def:I2D}
	We define the vector local interpolation function as 
	\begin{equation}\label{eq:GIF}
		\mathbf I(\a,s)=\big(I_x(\a,s)\ ,\ I_y(\a,s)\big)^\top
	\end{equation}
	where $I_{x,y}$ are the corresponding scalar global interpolation functions. 
\end{definition}	
After invoking (\ref{eq:I}) and the global point and $s-$derivative values $\Big\{v_{i,j}^{x,y},\; d_{i,j}^{x,y},\ i=1\dots N_\a,\; j=1\dots N_s\Big\}$, (\ref{eq:GIF}) reads as
\begin{align}
	\mathbf I(\a,s)=&\sum_{i=1}^{N_\a} \sum_{j=1}^{N_s} \(v_{i,j}^x\; ,\; v_{i,j}^y\)^\top  
						\(H_7^{C_{i-1,j-1}}+ H_5^{C_{i-1,j}} + H_3^{C_{i,j-1}} + H_1^{C_{i,j}}\)(\a,s) \nn\\
					&+\sum_{i=1}^{N_\a} \sum_{j=1}^{N_s}\(d_{i,j}^x\; ,\; d_{i,j}^y\)^\top  \(H_8^{C_{i-1,j-1}}+ H_6^{C_{i-1,j}} + H_4^{C_{i,j-1}} + H_2^{C_{i,j}}\)(\a,s)\;. \label{eq:I2D}
\end{align}
The corresponding \emph{2D global approximation space} reads:
\begin{equation}\label{eq:V2D}
	\mathcal V^{2d}=\left\{ \vec{v}:B\longrightarrow \R^2\  \Big|\  \vec{v} \big|_C \in \mathcal V_C^{2d}, \forall C \in \mathcal T_{\Delta \a, \Delta s}  \right\}\;.
\end{equation}

Based on the strong formulation (\ref{eq:strong}) and on (\ref{eq:V2D}) we present here, without further justification, the corresponding FE formulation for the numerical treatment of (\ref{eq:strong}) and refer to \cite{MOSS} for further details on the ``full'' FEM. The presentation is with respect to the family $\vec{F}$; symmetrically similar is the problem for the other filament family.
\begin{center}
\emph{
	Find continuous in time solutions $\vec{F}(\cdot,\cdot,t)\in \mathcal V^{2d}$ such that, for every test function $\vec{v}\in \mathcal V^{2d}$, 
	}
	\begin{align}
		0 	=&\int_{B}\eta \Big( \mu^B \pd^2_s\vec{F} \cdot \pd_s^2\vec{v} 
				+ \mu^A D_t \vec{F} \cdot \vec{v} 
				+ \lambda_\subtext{inext} \pd_s\vec{F} \cdot \pd_s\vec{v} \Big) \ d(\a,s) \nn\\
			&\mp \int_{B}  \eta\eta^*\Big( \mu^T_\pm(\phi-\phi_0)\pd_s\vec{F}^{\perp} \cdot \pd_s \vec{v} 
				\mp \mu^S \(D_t\vec{F} - D_t^*\vec{F}^* \)\cdot \vec{v}\Big) \ d(\a,s) \nn\\
			&-\int_{B} p(\varrho) \Big( \pd_\a \vec{F}^{\perp}\cdot \pd_s \vec{v} - \pd_s \vec{F}^{\perp}\cdot \pd_\a \vec{v}\Big) \ d(\a,s) \nn\\
			&+	\int_{\pd B \cap \{s=-L\}} \eta \Big( f_\subtext{tan}(\a)\pd_s\vec{F} + f_{\subtext{inn}}(\a)\vec{v}^\pm(\a) \Big)\cdot \vec{v}\ d\a \nn\\
			&\mp	\int_{\pd B \cap \{s=0\}} \lambda_\subtext{tether}\cdot \vec{v} \ d\a \;.\label{eq:FEM}
	\end{align}
\end{center}
	
	\subsection{Numerical considerations}

In this section we present some of the considerations that we took into account when addressing the numerical resolution of (\ref{eq:FEM}). In particular we address the numerical treatment of each of the terms of (\ref{eq:FEM}) separately.

Following (\ref{eq:IC2D}), $\vec{F}$ is represented in terms of the Lagrange-Hermite basis functions (\ref{eq:H}) over the computational cell $C\in\mathcal T_{\Delta \a,\Delta s}$ at the time step $t^k$ as
\begin{equation}\label{eq:lincomp}
	\vec{F}^{k}\Big|_{C} (\a, s) =  \left (\sum_{i=1}^{8}p^{k}_{i,x} H_i^{C} (\a, s),\ \sum_{i=1}^{8}p^{k}_{i,y} H_i^{C} (\a, s)\right)^\top\;,
\end{equation}
where $p^{k}_{i,x}$, $p^{k}_{i,y}$ $i=1\dots8$ denote the $x$ and $y$ respectively, point and derivative value DOFs assigned to the vertices of $C$ under lexicographic enumeration, see also (\ref{eq:verC}). There are some notable exceptions in this approach where it was deemed necessary to approximate the solution with the help of another basis set. We will address these terms with detail. As a rule though, for the derivation of the FEM, we replace (\ref{eq:lincomp}) in (\ref{eq:FEM}) and test against every shape function $H^C_k$ for $k=1\dots8$ and $C\in \mathcal T_{\Delta \a,\Delta s}$. 

\subsubsection{Filament length}
We compute the length of the filaments $L$, and the length distribution $\eta$ explicitly, following the modelling considerations described in \cite{MSOS} (see also \cite{Sfakianakis.2012}). The filament length $L$, in particular, is given by
\[L=-\frac{\kappa_\subtext{cap,eff}}{\kappa_\subtext{sev}}+\sqrt{\frac{\kappa_\subtext{cap,eff}^2}{\kappa_\subtext{sev}^2}
  +\frac{2v}{\kappa_\subtext{sev}} \log\left(\frac{\eta(s=0)}{\eta_\subtext{min}}\right)} \,,\]
where $\kappa_\subtext{sev}$, $\kappa_\subtext{cap,eff}$ are the severing and capping rates of the filaments, and where $\eta_\subtext{min}$ is a minimum cut-off density for F-actin. Note moreover that faster polymerization rate $v$ leads to wider lamellipodia.
\subsubsection{Filament bending}
The bending term 
reads in the FE formulation (\ref{eq:FEM}) as
\[
	\int_{C}\eta \pd^2_s\vec{F} \cdot \pd_s^2\vec{v}\  d(\a,s)\;.
\]

The numerical solution $F$ is disretized implicitly in time and is represented over the computational cell $C$ at the time step $t^{n+1}$ using (\ref{eq:lincomp}). It's contribution in the FEM formulation reads ---after testing against $H_k^{C}(\a,s)$, $k=1\dots8$--- in its $x$-coordinate as follows:
\[ 
	\eta \sum_{i=1}^{8}p^{n+1}_{i,x} \int_{C} \(\pd_s^2 H_i^{C}\,\pd_s^2 H_k^{C}\) (\a, s)\; d(\a,s)\;.
\]
\subsubsection{Adhesion with the substrate}
The adhesion term reads 
in the FE formulation (\ref{eq:FEM}) as
\[
	\int_{C} \eta D_t \vec{F} \cdot \vec{v} \ d(\a,s)\;,
\]
which reads after explicit in time discretization, as:
\[
	\int_{C} \(\frac{\vec{F}^{n+1}-\vec{F}^{n}}{\Delta t}-v\pd_s \vec{F}^{n} \)\cdot \vec{v} \ d(\a,s)\;.
\]
Expanding $\vec{F}^{n}$, $\vec{F}^{n+1}$ according to (\ref{eq:lincomp}) and testing against $H_k^{C}$,  it reads in the $x$-coordinate as 
\[
	\int_{C} \bigg( \frac{1}{\Delta t}\sum_{i=1}^{8} \( p^{n+1}_{i,x} - p^{n}_{i,x} \) H_i^{C}(\a,s) 
						-v \sum_{i=1}^{8} p^{n}_{i,x} \pd_s H_i^{C}(\a,s) \bigg) H_k^{C}(\a,s) \ d(\a,s)\;,
\]
or
\begin{eqnarray}
	\frac{1}{\Delta t}\sum_{i=1}^{8} p^{n+1}_{i,x} \int_{C} \(H_i^{C}\, H_k^{C}\)(\a,s) \ d(\a,s) 
	-\sum_	{i=1}^{8} p^{n}_{i,x} \bigg( \frac{1}{\Delta t}  \int_{C}\(H_i^{C}\,H_k^{C}\)(\a,s) \ d(\a,s) 
		+v  \int_{C}\(\pd_s H_i^{C}\, H_k^{C}\)(\a,s) \ d(\a,s) \bigg)\;.\nn
\end{eqnarray}
The implicit terms are added in the stiffness matrix, and the explicit ones in the right-hand side of the discrete FEM formulation.
\subsubsection{Stretching of cross-links}\label{sec:str}

The stretching of cross-links reads in the FE formulation (\ref{eq:FEM}) as
\begin{equation}\label{eq:str}
	\int_{C} \eta\eta^*\(D_t\vec{F} - D_t^*\vec{F}^* \)\cdot \vec{v} \ d(\a,s)\;.
\end{equation}
It involves filaments of both families and is derived under the assumption, that in a previous time $t-\delta t$ the filaments $\a$, $\a^*$ cross at the positions $s$, $s^*$.

Due to its nature, (\ref{eq:str}) necessitates special treatment: for every filament and discretization node $(\a,s)$ of one family, we assume that there exist $(\a^*,s^*)$ of the other family (in general non-discretization nodes) such that:
\begin{equation}
\label{eq:CrossAs}
    \vec{F}^{ n} (\a,s+ v\Delta t )=\vec{F}^{n,*}(\a^*,s^*+v^*\Delta t)\;,
\end{equation}
where $\vec{F}^{n}$ denotes the numerical solution of the $\vec{F}$ at time $t^n$. Accordingly, (\ref{eq:str}) reads, after time discretization, as:
\begin{align}
	\Delta t\big(D_t \vec{F} (\a,s ) -& D_t \vec{F}^* (\a^*,s^* )\big)\nn\\
		=&	\vec{F}^{n+1} (\a,s) 	- \vec{F}^{n} (\a,s )- v\Delta t \pd_s \vec{F}^{n} (\a,s ) \nn\\
			&- \vec{F}^{n+1,*} (\a^*,s^*) 	+ \vec{F}^{n,*} (\a^*,s^* )+ v^*\Delta t \pd_s \vec{F}^{n,*} (\a^*,s^* ) \nn\\
		=&	\vec{F}^{n+1} (\a,s) 	- \vec{F}^{n} (\a,s+ v\Delta t )\nn\\
			&- \vec{F}^{n+1,*} (\a^*,s^*) 	+ \vec{F}^{n,*} (\a^*,s^*+ v^*\Delta t)\nn\\
		=&	\vec{F}^{n+1} (\a,s )	- \vec{F}^{n+1,*} (\a^*,s^*)\;. \label{eq:Stretch}
\end{align}

The relation (\ref{eq:Stretch}), necessitates that for every $(\a,s)$-pair (discretization nodes) the corresponding $(\a^*,s^*)$-pair (decimal non-discretization nodes in general) should be computed. We do this in the following way: we approximate the crossing-point in terms of the $(*)$ family, by identifying the ``below'' and ``above'' (bounding) filaments and position nodes $\a_1^*,\a_2^*$ and $s_1^*, s_2^*$ respectively, see also Fig. \ref{fig:discr}. These provide with the four surrounding vertices of the crossing-point with respect to the $(*)$ family:
\begin{equation}\label{eq:vert}
	\left.
	\begin{array}{lcl}
	(q_{1x},q_{1y})^\top=\vec{F}^*(\a_1^*,s_1^*),&~~&  (q_{2x},q_{2y})^\top=\vec{F}^*(\a_1^*,s_2^*),\nn \\
	(q_{3x},q_{3y})^\top=\vec{F}^*(\a_2^*,s_1^*),& &  (q_{4x},q_{4y})^\top=\vec{F}^*(\a_2^*,s_2^*).
	\end{array}
	\right\}
\end{equation}
The corresponding point is approximated by the interpolation:
\begin{align}
	\vec{F}^*(\a^*,s^*)	&\approx (\overline{q_{x}},\overline{q_{y}})^\top \label{eq:inter}\\
					&= c_1 \vec{F}^*(\a_1^*,s_1^*)+c_2 \vec{F}^*(\a_1^*,s_2^*)
						+c_3 \vec{F}^*(\a_2^*,s_1^*)+c_4 \vec{F}^*(\a_2,s_2)\;, \label{eq:repr}
\end{align}
with the weights given through the integer parts $(\tilde \a,\tilde s)$ of $(\a^*,s^*)$:
\begin{equation}\label{eq:cs}
	\left.\begin{array}{lcl}
	c_1=(1-\tilde \a)(1-\tilde s), &~~& c_2=(1-\tilde \a)\tilde s,\\
	c_3=\tilde \a(1-\tilde s), && c_4=\tilde \a \tilde s.
	\end{array}\right\}
\end{equation}

Subsequently, we multiply with the Lagrange-Hermite shape functions $H_i(\alpha,s)$, $i=1,\dots,8$  (\ref{eq:H}) defined over the cells of the domain $B$ of $\vec{F}$. Since though the nodes of one family are represented only as decimal coordinates with respect to the other family,  one rectangular cell in $B$ corresponds to a quadrilateral in $B^*$ and covers there parts of different cells. This causes the following complication: if the ``usual'' Lagrange-Hermite elements are used for the $(*)$ family in the reconstruction of $\vec{F}^*$ over  $B$, it becomes unclear which direction should the derivatives take into account. It also turns out that the use of the more ``usual'' Lagrange-Hermite approximation for the $\vec{F}$ is problematic because the two parts are then implemented in an asymmetric way (the term acts of the points and the derivatives of $\vec{F}$, but only on the points of the $(*)$ family). We have used instead Lagrange-Lagrange\footnote{linear shape functions in both $\a$- and $s$-directions of the form $L_i^C(\a)\,L_j^C(s)$, $i,j=1,2$ with $L_\cdot^C(\cdot)$ given by (\ref{eq:LG})} elements for the approximation of both families.

Therefore, instead of using (\ref{eq:lincomp}), we expand $\vec{F}$ and $\vec{F}^*$ in the cell $C$ of $B$  as:
\begin{align}
	\vec{F}\Big|_{C} (\a, s) &=  \left (\sum_{i=1}^{4}p_{i,x}^{n+1} L_i^{C} (\a, s),\ \sum_{i=1}^{4}p_{i,y}^{n+1} L_i^{C} (\a, s)\right)^\top\;,\nn \\
	\vec{F}^*\Big|_{C} (\a, s) &=  \left (\sum_{i=1}^{4}\overline{q_{i,x}^{n+1}} L_i^{C} (\a, s),\ \sum_{i=1}^{4}\overline{q_{i,y}^{n+1}} L_i^{C} (\a, s)\right)^\top\;,\nn
\end{align}
where $p_{i,x}^{n+1}$, $p_{i,y}^{n+1}$, $\overline{q_{i,x}^{n+1}}$, $\overline{q_{i,y}^{n+1}}$ are the DOFs corresponding to the positions of the vertices of $C$ with respect to the two families. Due to the interpolation (\ref{eq:repr}) we write $\overline{q}_{i,x}^{n+1}=\sum_{r=1}^4 c_{i,r}^{n+1}q_{i,rx}^{n+1}$ and $\overline{q}_{i,y}^{n+1}=\sum_{r=1}^4 c_{i,r}^{n+1}q_{i,ry}^{n+1}$ with $q_{i,rx}^{n+1}$, $q_{i,ry}^{n+1}$, $c_{i,r}^{n+1}$, $r=1\dots4$ given by $(\ref{eq:vert})$ and (\ref{eq:cs}) respectively.

At the end, after testing against $H_k$, and integrating over $C$,  (\ref{eq:Stretch}) reads in the $x$-coordinate as:	
\[
	\int_{C}(\vec{F}^{n+1}- \vec{F}^{n+1,*}) H_k \longrightarrow \sum_{i=1}^4 \left(p_{i,x}^{n+1}-\sum_{r=1}^4 c_{i,r}^{n+1} q_{i,rx}^{n+1}\right)\int_{C} \(L_i^C\,H_k^C\)(\a,s)\;d(\a,s)\;.
\]

\subsubsection{Twisting of cross-links}
The twisting of the cross-links reads in the FE formulation (\ref{eq:FEM}) as
\[
	\int_{C}  \eta\eta^*(\phi-\phi_0)\pd_s\vec{F}^{\perp} \cdot \mathbf \pd_s \vec{v}\; d(\a,s)\;.
\]
The angle $\phi$ between the crossing filaments is approximated with a process similar to the one described in Sect. \ref{sec:str}. At every discretization node $(\a,s)$ of one family we identify the (probably decimal) node $(\a^*,s^*)$ of the other family such that $\vec{F}(\a,s)=\vec{F}^*(\a^*,s^*)$. We employ the interpolation formulas (\ref{eq:repr}) adjusted for the $\pd_s$: 
\begin{align}
\pd_s \vec{F}^*(\a^*,s^*)=& c_1 \pd_s \vec{F}^*(\a_1^*,s_1^*)+c_2 \pd_s \vec{F}^*(\a_1^*,s_2^*) \nn \\
							&+ c_3 \pd_s \vec{F}^*(\a_2^*,s_1^*)+c_4 \pd_s \vec{F}^*(\a_2^*,s_2^*)\;,
\end{align}
with weights $c_1\dots c_4$ given by (\ref{eq:cs}). Expanding now $\vec{F}\Big|_{C}$ implicitly (at $t^{n+1}$) according to (\ref{eq:lincomp}), and computing $\phi$ explicitly (at $t^n$) , the twisting term reads in its $x$-coordinate as
\[ 
	-\eta\eta^*\sum_{i=1}^{8}(\phi^n-\phi_0)p^{n+1}_{i,y} \int_{C} \(\pd_s H_i^{C}\, \pd_s H_k^{C}\) (\a, s)\; d(\a,s)\;.
\]
%
%
%

\subsubsection{Filament repulsion}
The pressure term reads in the FE formulation (\ref{eq:FEM}) as
\[
	\int_{C} p(\varrho) \( \pd_{\a}\vec{F}^{\perp} \cdot \pd_s \vec{v}
		\, -\,  \pd_s \vec{F}^{\perp} \cdot \pd_\a \vec{v} \)\; d(\a,s)
\]
where 
\begin{align}
	\varrho&=\frac{\eta}{| \pd_s \vec{F}\cdot\pd_\a\vec{F}^{\perp} |}\;, \label{eq:dens}\\
	p(\varrho)&=\mu^P \varrho\;. \label{eq:pres}
\end{align}
For the computation of $\varrho$, and in effect of $p(\varrho)$, we approximate the point values $\vec{F} (\a,s)\Big|_{C}$ in (\ref{eq:dens}), by the ---explicit in time--- cell averages 
\begin{equation}\label{eq:appr}
	\frac{1}{\Delta s \Delta \a} \int_{C} \vec{F}^{n}(\a,s)\; d(\a,s) = \frac{1}{\Delta s \Delta \a} \sum_{i=1}^{8} (p_{i,x}^{n},p_{i,y}^{n})^\top \int_{C} H_i^{C}(\a,s)\; d(\a,s)\;. 
\end{equation}
After implicit discretization and expansion of $F\Big|_{C}$ according to (\ref{eq:lincomp}), the pressure term reads in the $x$-coordinate as:
\[
	-\sum_{i=1}^8 p_{i,y}^{n+1} p(\varrho)\(\int_{C} \(\pd_\a H_i^{C}\, \pd_sH_k^{C}\)(\a,s)\; d(\a,s) +\int_{C} \(\pd_s H_i^{C}\, \pd_\a H_k^{C}\)(\a,s)\; d(\a,s) \)\;,
\]
with $p(\varrho)$ given by (\ref{eq:pres}), (\ref{eq:dens}), (\ref{eq:appr}).


\subsubsection{Innerforces}
The computation of the inner pulling forces $f_\subtext{tan}(\a)$ and $f_\subtext{inn}(\a)$ follow from the force balance law (see also the BCs (\ref{eq:newBC}) at $s=-L$):
\begin{eqnarray}\label{eq:constraint}
	\int_{[0,2\pi)} \eta(\a,s) \Big( f_\subtext{tan}(\a) \pd_s \vec{F}(\a,s) +f_\subtext{inn}(\a) \vec{V}(\a) \Big)\bigg |_{s=-L} {d}\a=0\;.
\end{eqnarray}
In effect, this leads in a variational approach to the minimization of 
\[
	\int_{[0,2\pi)} \eta(\a,s) \Big( \frac{1}{\gamma}\(f_\subtext{tan}(\a) -\gamma A\)^2 +\frac{1}{1-\gamma}\( f_\subtext{inn}(\a) -(1-\gamma)A\)^2 \Big) \bigg|_{s=-L}d\a
\]
under the constraint (\ref{eq:constraint}), where $A=\mu^\subtext{IP}(A_c-A_0)_+$ measures the positive deviation of the area $A_c$ occupied by the cell from an equilibrium value $A_0$. The constraint minimization problem yields
\begin{align}
	f_\subtext{tan}(\a)  &=  \gamma A\( 1- \(\mu_1, \mu_2\)^\top \cdot \pd_s \vec{F}(\a,s)\bigg|_{s=-L} \)\;,\label{eq:f_tan}\\
	f_\subtext{inn}(\a)  &=  1-\gamma A\( 1- \(\mu_1, \mu_2\)^\top \cdot \vec{V}(\a) \)\;,\label{eq:f_inn}
\end{align}
where
\begin{align}
	\(\mu_1, \mu_2\)^\top  =  &\(\int_{[0,2\pi)} \eta\( \gamma \pd_s \vec{F} \otimes \pd_s \vec{F} +(1-\gamma)\vec{V}(\a) \otimes \vec{V}(\a) \)\bigg|_{s=-L} {d}\a\)^{-1} \nn \\
	&\times\int_{[0,2\pi)} \eta \( \gamma\pd_s \vec{F} + (1-\gamma) \vec{V}(\a)\)\bigg|_{s=-L} {d}\a\;, \label{eq:mu1mu2}
\end{align}
where $(\times)$ stands for the regular number multiplication.

Using (\ref{eq:f_tan}) and $s_0=-L$, we treat the first term of (\ref{eq:constraint}) as follows (the second term is treated similarly)
\[
	\int_{[0,2\pi)} \eta(\a,s_0)\gamma A \( \pd_s \vec{F}(\a,s_0) - (\mu_1,\mu_2)^\top\cdot \pd_s \vec{F}(\a,s_0) \pd_s \vec{F}(\a,s_0)\)\; d\a\;,
\]
or, after expanding $\vec{F}\Big|_{C}$ explicitly at $t^{n}$ according to (\ref{eq:lincomp}), testing against $H_k$, it reads in the $x$-coordinate as
\begin{align}
	\eta\gamma A\(\sum_{i=1,..,8} p_{i,x}^{n} \int \(\pd_s H_i\, H_k\) (\a,s_0)\; d\a
	-\sum_{i,j=1}^8 p_{i,x}^{n} \( \(p_{i,x}^{n}, p_{i,y}^{n}\)^\top\cdot (\mu_1,\mu_2)^\top\) 
		\int \(\pd_s H_i\,\pd_s H_j\, H_k\)(\a,s_0)\; d\a\)\;,\nn
\end{align}
where the integrations are over each inner element of the discretization and where $(\mu_1,\mu_2)^\top$ is computed explicitly in time via (\ref{eq:mu1mu2}).

\subsubsection{In-extensibility}\label{sec:inext}
The in-extensibility term reads in the FEM (\ref{eq:FEM}) as
\[
	\int_{C}\eta \lambda_\subtext{inext}\pd_s \vec{F} \cdot \pd_s \vec{v}\; d(\a,s)\;.
\]
We employ the Augmented Lagrangian approach to evaluate $\lambda_\subtext{inext}$; the strong formulation of which recasts into 
\[
	\pd_s\(\eta \( \lambda^{n} +  \frac{1}{\epsilon} \( \pd_s \vec{F}^{n} \cdot \pd_s \vec{F}^{n+1}-1\) \) \pd_s \vec{F}^{n}\)\;,
\]
with $\lambda^{n}$ been updated in every time step by
\[
	\lambda^{n+1}=\lambda^{n}+\frac{1}{\epsilon}\left(\pd_s \vec{F}^{n} \cdot \pd_s \vec{F}^{n+1}-1\right)\;,
\]
for $0<\epsilon<1$.
When linearizing as $\pd_s \vec{F} \longmapsto \left(|\pd_s \vec{F}|^2-1\right)\pd_s \vec{F}$ at $\pd_s \vec{F}^n$, the punishing term $-\frac{1}{\epsilon}\pd_s \Big(\left(|\pd_s \mathbf  F|^2-1\right)\pd_s \vec{F}\Big)$ is approximated by
\[
	-\frac{1}{\epsilon}\pd_s\Big( \left(|\pd_s \vec{F}^n|^2-1\right)\pd_s \vec{F}^{n+1}+2\left(\pd_s \vec{F}^n \cdot \pd_s \vec{F}^{n+1}-|\pd_s \vec{F}^n|^2\right)\pd_s \vec{F}^n \Big)\;,
\]
revealing a contribution in the $\pd_s$-direction of both the old and the new time step.
	
In effect, the in-extensibility term reads as:
\begin{equation}\label{eq:z_inext}
	-\pd_s\left( \Big(\lambda^{n}+\frac{1}{\epsilon}\left(|\pd_s \vec{F}^{n}|^2-1\right)\Big)\pd_s \vec{F}^{n+1}+\frac{2}{\epsilon}\left(\pd_s \vec{F}^{n} \cdot \pd_s \vec{F}^{n+1}-|\pd_s \vec{F}^{n}|^2\right)\pd_s \vec{F}^{n} \right)\;,
\end{equation}
with $\lambda^{n}$ given by
\begin{equation}\label{eq:l_inext}
	\lambda^{n+1}=\lambda^{n}+\frac{1}{\epsilon}\big( |\pd_s \vec{F}^{n}|^2 - 1 \big) + \frac{2}{\epsilon}\big(\pd_s \vec{F}^{n} \cdot \pd_s \vec{F}^{n+1} - |\pd_s \vec{F}^{n}|^2 \big)\;.
\end{equation}


In a way similar to the previous terms, we expand $\vec{F}^n\Big|_{C}$, $\vec{F}^{n+1}\Big|_{C}$ according to (\ref{eq:lincomp}), $\lambda^n$ by $\sum_{r=1}^4 \bar{\lambda}_r^n $, and we test against  $H_k$ to obtain the final form of the in-extensibility term (\ref{eq:z_inext}). The formula contains both explicit and implicit contributions and is omitted for the sake of the presentation.

\subsection{Reparametrization}\label{sec:repar}

The direct implementation of the FE formulation (\ref{eq:FEM}) as was described in the previous paragraphs has been seen to have two drawbacks that we have addressed with proper reparametrizations, at every time step, along the $\a$- and $s$-directions separately. We sketch here the suggested reparametrization and the treatment that we have provided and refer to \cite{MOSS} for further detail.

\subsubsection{In $\a$-direction}
We reparametrize along the $\a$-direction to ensure that the ``computational'' filaments (mappings of the discretizations nodes along the $\a$-direction) are ``regularly'' distributed over the physical space $\Omega(t)$ (see Fig. \ref{fig:domains}). In particular, we define a mapping $g(\cdot,t):[0,2\pi) \rightarrow [0,2\pi)$, with $\beta \mapsto\a:=g(\beta,t)$, which leads to a new weak formulation of the problem. The original bending and adhesion terms, for example: 
\begin{eqnarray}
	\mu_B \pd_s^2(\eta \pd_s^2 \mathbf F) + \mu^A \eta D_t \mathbf F\nn
\end{eqnarray}
recast into:
\begin{eqnarray}
	&& \mu_B \pd_s^2(\tilde{\eta} \pd_s^2 \tilde{\vec F}) + \mu^A \tilde{ \eta} D_t^{\beta} \tilde{ \vec F }\nn
\end{eqnarray}
where $\mathbf F(\a,s) = \tilde{ \mathbf F}(\beta,s)$, and 
\begin{align}
	\tilde{ \eta} (\beta,s) &= \eta(\a,s)\left|g'(\beta)\right|\;,	\label{eq:etadash} \\
	D_t^{\beta} &= \pd_t - v\pd_s\;, \label{eq:Dtbeta}
\end{align}
where (\ref{eq:Dtbeta}) is derived under the simplification assumption that $\dot{g}^{-1}\approx 0$. Note also that, in (\ref{eq:etadash}), we have incorporated the contribution of $g$ into $\tilde{ \eta}$. 

One way to define such $g$ that maintains the ``regular'' distribution of the ``computational'' filaments is the following:
\[\a \longmapsto g^{-1}(\a,t) =:\beta=\frac{\int_0^\a \left|\pd_{\a}\mathbf F(\tilde{\a},0,t)\right|{d}\tilde{\a}}{\int_0^{2\pi} \left|\pd_{\a}\mathbf F(\tilde{\a},0,t)\right|{d}\tilde{\a}} 2\pi\;,\]
which for $M(t)=\int_0^{2\pi} \left|\pd_{\a}F(\tilde{\a},0,t)\right|{d}\tilde{\a}$ reads $|\pd_{\beta}\tilde{F}(\beta,0,t)|=\frac{M(t)}{2\pi}$. 

Numerically, this is achieved by setting $L_i=\frac{\subtext{length of $i$-th outer segment}}{\subtext{total length of membrane}}\,2\pi$, 
and defining the piecewise linear $g$:
\begin{eqnarray}
	g:\left[\sum_{j=1}^{i-1}L_j, \sum_{j=1}^{i}L_j\right]&\longrightarrow& [(i-1)\,\Delta\a,i\,\Delta\a]\nn\\
	\beta &\longmapsto& \a:= \Delta \a(i-1)+\Delta \a\frac{\beta-\sum_{j=1}^{i-1}L_j}{L_i}\nn
\end{eqnarray}
where $\Delta \a$ is the discretization step size along the  $\a$-direction.

\subsubsection{In $s$-direction}
Uneven filaments lengths, lead to discretizations with uneven $\Delta s$ in the opposite sides $\a_1$, $\a_2$ of the cell $C$, see (\ref{eq:C}). In some cases this can lead to numerical instabilities in the form of filament oscillations. To address this issue we have remapped all filaments to a constant length and transferred the control of the filament length to the in-extensibility terms. 

In particular, we set a map from $[-L(\a,t),0]$ onto $[-L,0]$ for a constant $L$ via the change of variables:
\begin{eqnarray}
	(s,\a) &\rightarrow& \(\frac{s}{ L(\a)}, \a\) \nn
\end{eqnarray}
where $L(\a)$ is the ratio between the original constant length $L$ and the new length. Accordingly, the strong formulation of the problem (\ref{eq:strong}) is transformed to read
\begin{align}
	\pd_s^2\(\eta\pd_s^2\mathbf F\)&+(L(\a))^4\eta\(\pd_t-\frac{v}{L(\a)}\pd_s\)\mathbf F\nn \\
	&+(L(\a))^2\pd_s\(\eta\eta^*(\phi-\phi_0)\pd_s \mathbf F^{\perp}\)\nn \\
	&+\eta\eta^*(L(\a))^4\(\(\pd_t-\frac{v}{L(\a)}\pd_s\)\mathbf F
		-\(\pd_t-\frac{v^{*}}{L^{*}(\a)}\pd_s\)\mathbf F^*\)\nn \\
	&- \pd_s\(\eta \lambda_\subtext{inext}  \pd_s \mathbf F\)\nn \\
	&+(L(\a))^3\(\pd_s \(\rho\pd_{\a}\mathbf F^{\perp}\)-\pd_\a \(\rho\pd_s \mathbf F^\perp\)\)=0\;. \label{eq:strong_trans}
\end{align}
where the (punishing) in-extensibility $\lambda_\subtext{inext}$ now reads 
\[
	\lambda_\subtext{inext}={\frac{1}{\epsilon} \(|\pd_s \mathbf F|^2-(L(\a))^2\)},
\]
accounting hence for the proper length of the filaments. This term is treated numerically in the way described in Sect. \ref{sec:inext}. The $\frac{1}{L(\a)}$ appearing in the stretching and adhesion terms is absorbed by redefining the polymerization velocities. Note also that $L(\a)$ does not depend on $s$, hence all integrations by parts in the $s$-direction are not affected. 
	
\section{Numerical simulations}
The purpose of this section is to demonstrate that the model is capable of describing complex biological experiments. In \cite{MSOS} several numerical experiments were described which demonstrated the effects of the the individual terms of the model. Additionally we demonstrated that the model can be used to simulate chemotaxis and to study how different signaling processes affect cell shape and filament density. Here we go a step further and simulate how the shape of a migrating cell is influenced by spatially selective adhesion patterns. Such studies are used to better understand the force balance between adhesions, contraction mechanisms, actin polymerization and other network proteins.

\subsection{Experiment 1: Adhesive{\textbackslash}Less-adhesive stripes}
\begin{figure}[t]
        \centering       
                \includegraphics[width=0.9\textwidth]{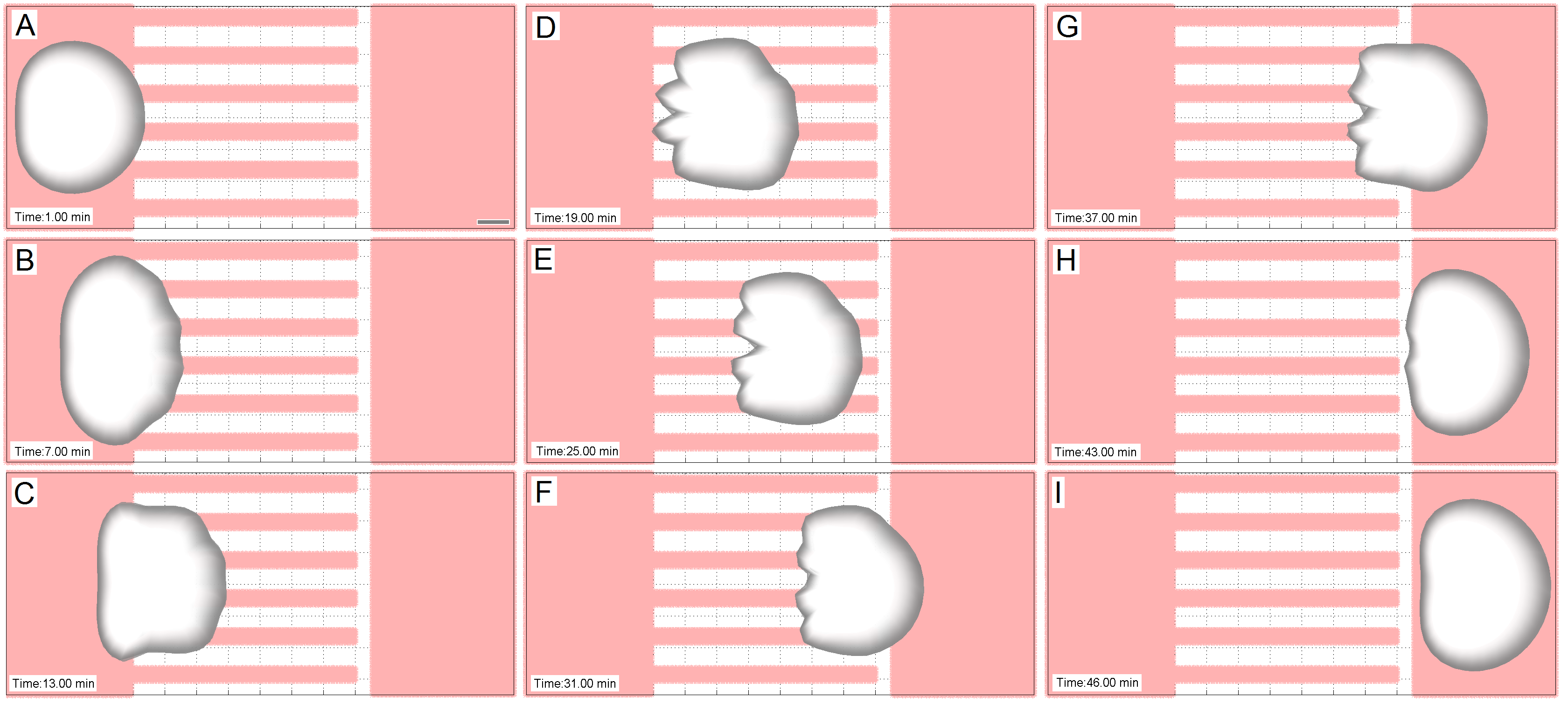}
        \caption{\small Movement of a cell on an adhesive substrate (red) with less-adhesive stripes (white). Shading represents actin network density. A-I: Times series of a cell moving over a stripe pattern ($80\%$ drop in adhesiveness). Parameter values as in Table \ref{tab:parameters}. The bar represents $\rm 10\, \umu m$.}
	\label{fig:stripes}
\end{figure}

\begin{figure}[t]
        \centering       
                \includegraphics[width=0.7\textwidth]{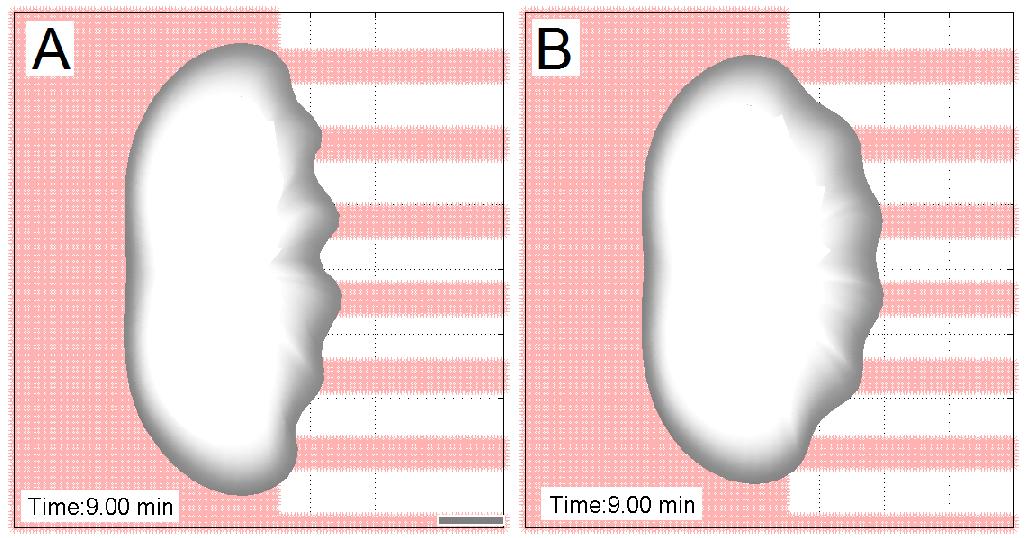}
        \caption{\small Movement of a cell on an adhesive substrate (red) with less-adhesive stripes (white). Shading represents actin network density. A: $90\%$ drop in adhesiveness, B: $80\%$ drop in adhesiveness. Parameter values as in Table \ref{tab:parameters}. The bar represents $\rm 10\, \umu m$.}
	\label{fig:stripes_fine}
\end{figure}


In this series of experiments we show that the model's behavior is similar to that of cells in the experiments described in \cite{Csucs2007}. In these experiments migrating fish keratocytes were placed on substrates which were prepared to have chemical patterns of the Extracellular Matrix (ECM) protein, fibronectin. This protein is a ligand of integrin, an important component of adhesions. In \cite{Csucs2007} stripe pattern were used with $\rm 5\,\umu m$ wide adhesive (fibronectin containing) stripes and non-adhesive stripes (without fibronectin) with widths varying between $\rm 5\, \umu m$ and $\rm 30\, \umu m$. In \cite{Csucs2007} it was described, that cell shape was affected in a very distinct way: protruding bumps on the adhesive stripes and lagging bumps on the non-adhesive stripes were observed and their width was correlated to the stripe width. Also it was observed that cells tend to symmetrize themselves such that they had an equal number of adhesive stripes to the right and to the left of their cell center. 
In the numerical experiment we used the same geometrical pattern with $\rm 5\,\umu m$ adhesive stripes interspaced with $\rm 7\,\umu m $ less-adhesive stripes. In the mathematical model, the adhesion forces result in friction with the ground and, speaking in numerical terms, they link one time step with the next. Therefore we cannot locally set the adhesion coefficient to zero, since this would render the stiffness matrix degenerate. Hence the adhesion coefficient in the less-adhesive regions was decreased to $10-20\%$ as compared to the adhesive regions. Whilst the keratocytes in the experiments by \cite{Csucs2007} move spontaneously without an external signal, we simulate chemotactic cells, since at this point the model cannot describe the dynamics of contractile bundles observed in keratocytes. However the numerical results show, that there are many similarities as far as general behavior and morphology are concerned, suggesting that the underlying phenomena are very similar. In Fig. \ref{fig:stripes}A--I a time series of the behavior of a cell on stripes with a drop in adhesiveness of $80\%$ is depicted. The following agreements between the simulation and the experiments were found: 
\begin{itemize}
\item On the striped region the cell shaped became more rectangular as compared the the crescent shape in the homogeneously adhesive region.
\item The numerical cell shows protruding bumps on the adhesive stripes and lagging bumps on non-adhesive stripes.
\item The simulation started with the cell being slightly non-symmetric with respect to the adhesive stripes and changed its position to have an equal number of adhesive stripes to the right and to the left of its center.
\item Spikes at the cell rear were observed.
\item After leaving the striped region the cell resumed its crescent shape and continued to migrate as before.
\end{itemize}
In Fig. \ref{fig:stripes_fine}A and B a comparison between the bumps on stripes with a $90\%$ (A) and a $80\%$ (B) drop in adhesiveness is shown. Here the $\a$-discretization used was twice as fine to allow for better analysis. As expected the bumps of the $90\%$-drop stripes are more pronounced. Over a time interval of several minutes we also observed the fluctuations in bump width observed in \cite{Csucs2007}.

\subsection{Experiment 2: Less-adhesive spikes on strongly adhesive ground}

\begin{figure}[t]
        \centering       
                \includegraphics[width=0.9\textwidth]{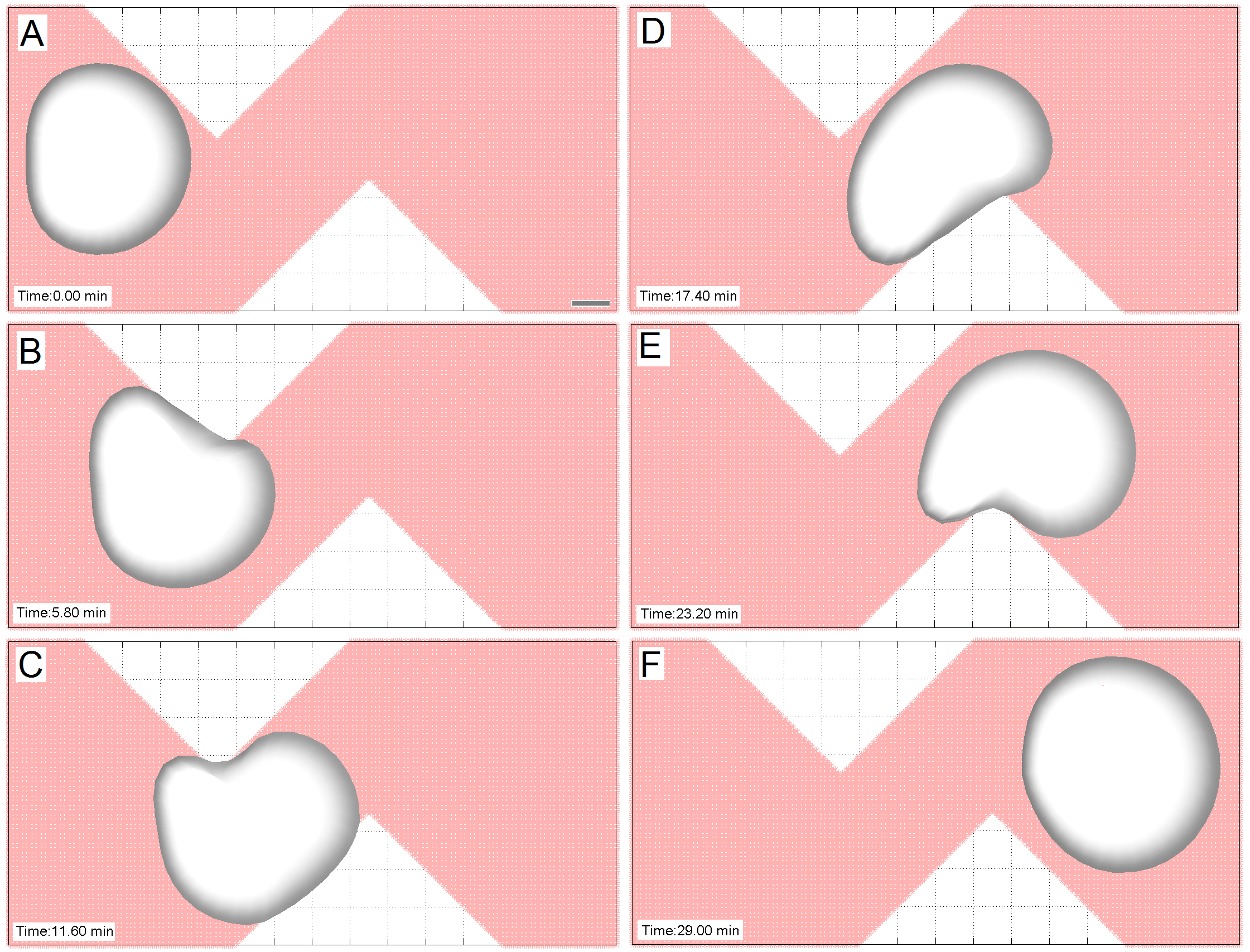}
        \caption{\small Movement of a cell on an adhesive substrate (red) with less-adhesive spikes (white). Shading represents actin network density. Parameter values as in Table \ref{tab:parameters}. The bar represents $\rm 10\, \umu m$.}
	\label{fig:spikes}
\end{figure}

In the next simulation we describe the behavior of a migrating cell on a substrate with a different pattern, showing that the model can make predictions for future biological experiments. We use the same setup as above with a pattern which consists of two shifted less-adhesive spikes from above and below. The drop in adhesiveness was chosen to be $80\%$. As opposed to the situation above, the cell is now able to almost fully avoid the less-adhesive regions. The behavior observed over a time of $\rm 30\, min$ is depicted in the time series in Fig. \ref{fig:spikes}. It can be seen that the cell behaves as if the less-adhesive spikes were obstacles and always only a very small fraction the lamellipodial region enters the less-adhesive areas.

\subsection{Parameters values:}
For the discretization we used a time step of $0.002$ and nine nodes per filament. For the first experiment we used 36 and 72 discrete filaments, for the second one  36.

For the biological parameters, we used the same as those in \cite{MSOS}, apart from the adhesion coefficient which was increased for the adhesive regions and decreased for the less-adhesive regions. They are summarized in Table \ref{tab:parameters}.

\begin{table}[t]
\caption{Parameter Values}
\label{tab:parameters}       
\begin{tabular}{| c | p{0.30\linewidth} | p{0.31\linewidth} | p{0.30\linewidth} | }
\hline
Var. & Meaning & Value & Comment \\ 
\hline
& & & \\
$\umu^B$ & bending elasticity & $\rm 0.07\,pN\,\umu m^2$ &\cite{Gittes1993}  \\
$\umu^A$ & macroscopic friction caused by adhesions & $\rm 0.041, 0.082, 0.41\,pN\, min\, \umu m^{-2}$ & lower values for less-adhesive regions, highest value for adhesive regions, order of magnitude from measurements in \cite{Li2003,Oberhauser2002}, estimation and calculations in \cite{Oelz2008,Oelz2010a,Schmeiser2010}\\
$\kappa_{br}$ & branching rate & $\rm 10\, min^{-1} $ & order of magnitude from \cite{Grimm2003}, chosen to fit $2 \overline \rho_\subtext{ref}=\rm 90\, \umu m^{-1}$ \cite{Small2002}\\
$\kappa_{cap}$ & capping rate & $\rm 5\, min^{-1} $ & order of magnitude from \cite{Grimm2003}, chosen to fit $2 \overline \rho_\subtext{ref}=\rm 90\, \umu m^{-1}$ \cite{Small2002}\\
$c_{rec}$ & Arp$2/3$ recruitment  & $\rm 900\, \umu m^{-1}\, min^{-1} $ & chosen to fit $2 \overline \rho_\subtext{ref}=\rm 90\, \umu m^{-1}$ \cite{Small2002}\\
$\kappa_{sev}$ & severing rate & $\rm 0.38\, min^{-1}\, \umu m^{-1}$ & chosen to give lamellipodium widths similar as described in \cite{Small2002} \\
$\umu^{IP}$ & actin-myosin interaction strength& $\rm 0.1\, pN\, \umu m^{-2}$ &  \\
$A_0$ & equilibrium inner area & $\rm 450\, \umu m^2$ & order of magnitude as in \cite{Verkhovsky1999,Small1978}\\
$v_\subtext{min}$ & minimal polymerization speed & $\rm 1.5\,\umu m/min^{-1}$ & in biological range\\
$v_\subtext{max}$ & maximal polymerization speed & $\rm 8\,\umu m/min^{-1}$ & in biological range\\
$\umu^P $ &pressure constant & $\rm 0.05\, pN\, \umu m $ & \\
$\umu^S$ & cross-link stretching constant & $\rm 7.1\!\times\! 10^{-3}\, pN\, min\, \umu m^{-1} $& \\
$\umu^T$ & cross-link twisting constant & $\rm 7.1\times 10^{-3}\, \umu m$ &\\
$\kappa_{ref}$ & reference leading edge curvature for polymerization speed reduction & $\rm (5\,\umu m)^{-1}$ & \\
\hline
\end{tabular}
\end{table}



\textbf{Aknowledgements:}
N.Sfakianakis wishes to thank the Alexander von Humboldt Foundation and the Center of Computational Sciences (CSM) of Mainz for their support, and M. Lukacova for the fruitful discussions during the preparation of this manuscript.
\bibliographystyle{plain} 
\bibliography{Basic}
\end{document}